\documentclass[useAMS,usenatbib]{mn2e}
\usepackage[normalem]{ulem}

\usepackage{graphicx}
\usepackage{times}

\usepackage{amsmath}
\usepackage{amssymb}
\usepackage{tabularx}
\usepackage{txfonts}
\usepackage{wasysym}
\usepackage{ragged2e}
\usepackage{natbib}
\usepackage[english]{babel}
\usepackage{float}

\usepackage{color}

\usepackage{aas}

\title[$\Delta$Y and $\Delta$AGE of MSPs in NGC\,6352]{The {\it Hubble Space Telescope} UV Legacy Survey of Galactic
 Globular Clusters. IV. Helium content and relative age of multiple stellar
  populations within NGC\,6352
          \thanks{           Based on observations with  the
                               NASA/ESA {\it Hubble Space Telescope},
                               obtained at  the Space Telescope Science
                               Institute,  which is operated by AURA, Inc.,
                               under NASA contract NAS 5-26555 under GO~13297.}}
\author[Nardiello et al.]{
D.\,Nardiello$^{1,2,3}$\thanks{E-mail: domenico.nardiello@studenti.unipd.it\,.}, 
G.\,Piotto$^{1,2}$, 
A.\,P.\, Milone$^{3}$,
A.\,F.\, Marino$^{3}$, 
L.\,R.\, Bedin$^{2}$,
J.\,Anderson$^{4}$,
\newauthor
A.\,Aparicio$^{5,6}$,
A.\,Bellini$^{4}$,
S.\,Cassisi$^{7}$,
F.\,D'Antona$^{8}$,
S.\,Hidalgo$^{5,6}$,
S.\,Ortolani$^{1,2}$,
\newauthor
A.\,Pietrinferni$^{7}$,
A.\,Renzini$^{2}$,
M.\,Salaris$^{9}$,
R.\,P.\,van der Marel$^{4}$,
and E.\,Vesperini$^{10}$
\\
$^{1}$Dipartimento di Fisica e Astronomia ``Galileo Galilei'', Universit\`a di Padova, Vicolo dell'Osservatorio 3, Padova IT-35122 \\
$^{2}$Istituto Nazionale di Astrofisica - Osservatorio Astronomico di Padova, Vicolo dell'Osservatorio 5, Padova, IT-35122 \\
$^{3}$Research School of Astronomy and Astrophysics, The Australian National University, Cotter Road, Weston, ACT, 2611, Australia \\
$^{4}$Space Telescope Science Institute, 3800 San Martin Drive, Baltimore, MD 21218, USA \\
$^{5}$Instituto de Astrof\`isica de Canarias, E-38200 La Laguna, Tenerife, Canary Islands, Spain \\
$^{6}$Department of Astrophysics, University of La Laguna, E-38200 La Laguna, Tenerife, Canary Islands, Spain\\
$^{7}$Istituto Nazionale di Astrofisica - Osservatorio Astronomico di Teramo, Via Mentore Maggini s.n.c., I-64100 Teramo, Italy \\
$^{8}$Istituto Nazionale di Astrofisica - Osservatorio Astronomico di Roma, Via Frascati 33, I-00040 Monteporzio Catone, Roma, Italy \\
$^{9}$Astrophysics Research Institute, Liverpool John Moores University, Liverpool Science Park, IC2 Building, 146 Brownlow Hill, Liverpool L3 5RF, UK\\
$^{10}$Department of Astronomy, Indiana University, Bloomington, IN 47405, USA
}

\begin{document}

\date{Accepted 2015 April 29; Received 2015 April 29; in original form 2015 April 15}

\pagerange{\pageref{firstpage}--\pageref{lastpage}} \pubyear{2002}

\maketitle

\label{firstpage}

\begin{abstract}
In this paper we combine WFC3/UVIS F275W, F336W, and F438W data from
the ``UV Legacy Survey of Galactic Globular Clusters: Shedding Light
on Their Populations and Formation'' (GO~13297) {\it HST} Treasury
program with F606W, F625W, F658N, and F814W ACS archive data for a
multi-wavelength study of the globular cluster NGC\,6352. In the
color-magnitude and two-color diagrams obtained with appropriate
combination of the photometry in the different bands we separate two
distinct stellar populations and trace them from the main sequence to
the subgiant, red giant, horizontal and asymptotic giant branches.

We infer that the two populations differ in He by $\Delta
Y=0.029\pm0.006$. With a new method, we also estimate the age
difference between the two sequences. Assuming no difference in [Fe/H]
and [$\alpha$/Fe], and the uncertainties on $\Delta Y$, we found a
difference in age between the two populations of $10 \pm 120$\,Myr.
If we assume [Fe/H] and [$\alpha$/Fe] differences of 0.02 dex (well
within the uncertainties of spectroscopic measurements), the total
uncertainty in the relative age rises to $\sim$300~Myr.
\end{abstract}

\begin{keywords}
stars: Population II --- globular clusters individual:\,NGC\,6352
\end{keywords}

\section{Introduction}
\label{sec:introduction}

The {\it Hubble Space Telescope} ({\it HST}) ``UV Legacy Survey of
Galactic Globular Clusters: Shedding Light on Their Populations and
Formation'' (GO~13297, PI Piotto) is a Treasury program to study
globular cluster (GC) stellar populations using multi-wavelength
high-precision {\it HST} photometry and astrometry. The program is
fully described in \citet[hereafter Paper\,I]{2015AJ....149...91P}.
The main purpose of this survey is to identify multiple stellar
populations and study their properties, including their relative ages,
chemical compositions, spatial distributions, and kinematics.  Within
GO~13297 we collected F275W, F336W, and F438W WFC3/UVIS images,
approximately overlapping the F606W and F814W data from GO~10755 (PI
Sarajedini, see \citealt{2007AJ....133.1658S}).
%%%%%% 

In this paper we analyze photometry of stars in the Galactic GC
NGC\,6352. Our main purpose is to infer the helium content and
determine the relative ages of its stellar populations.
NGC\,6352 is a Bulge (Galactic coordinates: $l=341$, $b=-7.2$),
metal-rich GC, located at 5.6 Kpc from the Sun [$(m-M)_V$=14.43,
  $E(B-V)$=0.22, \citealt[2010 revision]{1996AJ....112.1487H}], 
  and having luminosity $M_{V}=-6.47$ (\citealt[2010
    revision]{1996AJ....112.1487H}) and mass of
  $3.7\times10^4\,M_{\odot}$ (\citealt{2010MNRAS.406.2000M}).  From
high-resolution spectroscopy of nine horizontal branch (HB) stars,
\citet{2009A&A...493..913F} confirmed the results of previous works, that this cluster has a high
metallicity ([Fe/H]$\sim -0.55$) and is $\alpha-$enhanced
([$\alpha$/Fe]$\sim$0.2).

The paper is organized as follows. Section 2 presents the data and
data analysis. Section 3 and 4 show the characteristics of the two
stellar populations hosted by NGC\,6352, and their properties as seen
with different combinations of colors and magnitudes. In Section 5 the
difference in the helium content of the two stellar populations is
calculated.  In Section 6 we describe the new method we developed to
estimate the difference in age between the two stellar populations. In
Section 7 there are the conclusions.

%%%%%%%%%%%%%%%%%%%%%%%%%%%%%%%%%%%%%%%%%%%%%%%%%%%%%%%%%%%%%%%%%%%%%%%%%%%%%%%%%%%%

\section{Observations and data analysis}
\label{observations}
In order to identify the multiple stellar populations in NGC\,6352 we
used WFC3/UVIS images of GO~13297 and ACS/WFC data of GO~10775.  The
WFC3 data set consists in $2\times$706\,s +$2\times800$\,s F275W,
$4\times311$\,s F336W, and a 58\,s + 72\,s F438W images.  ACS data
include $4\times140$\,s + 7\,s F606W and $4\times150$\,s + 7\,s F814W
images, overlapped to WFC3 images.  We also reduced ACS images
collected in F625W and F658N bands within GO~12746 (PI Kong). Exposure
times are 2$\times$150\,s in F625W and a 650\,s + 643\,s in F658N.  A
detailed description of the reduction of GO~13297 data is provided in
Paper\,I (see their Sect.\,5).  We have used the photometric and
astrometric catalogs published by \citet{2008AJ....135.2114A} for
GO~10775 data, while photometry and astrometry of archive ACS/WFC
images from GO~12746 have been carried out as in
\citet{2008AJ....135.2114A}.

%%%%%%%%%%%%%%%%%%%%%%%
\section{The color-magnitude diagrams of NGC\,6352}
\label{sec:multipop}
\begin{figure}
  \centering
  \includegraphics[width=0.95\hsize]{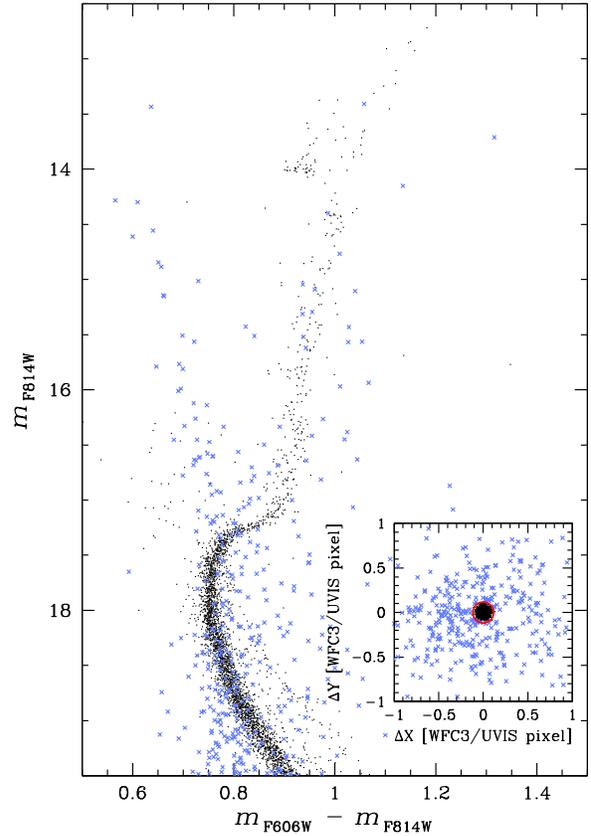}
  \caption{The $m_{\rm F814W}$ vs.  $m_{\rm F606W}-m_{\rm F814W}$ diagram
    of NGC\,6352. In black and in light blue are plotted cluster members and
    field stars, respectively.  Cluster members have been selected on
    the basis of their proper motions. The vector-points diagram of
    proper motions is plotted in the inset. 
    Black points within red circle and light blue points outside red circle
    indicate members and field stars, respectively.
  }
  \label{pm}
\end{figure}

In Paper\,I (see their Fig.~2), we have shown that the color-magnitude
diagram (CMD) of NGC\,6352 is not consistent with a single stellar
population.  The $m_{\rm F814W}$ vs.  $m_{\rm F606W}-m_{\rm F814W}$
CMD of the cluster members of NGC\,6352 is shown in Fig.~\ref{pm}: in
black there are the stars that, on the basis of their proper motions,
have high probability to be cluster members, in  light
  blue the rejected stars. 
 To derive stellar proper motions, we determined the displacement
  between the stellar positions in GO~10775 data-set (2006.4) and in
  GO~13297 data-set (2013.7 and 2014.5) by following the method described in
  detail by \citet[see their Sect.~4]{2012ApJ...760...39P}. The maximum time
  baseline for the proper motion measurements is 8.1 yrs.

\begin{figure*}
  \centering
  \includegraphics[width=0.95\hsize]{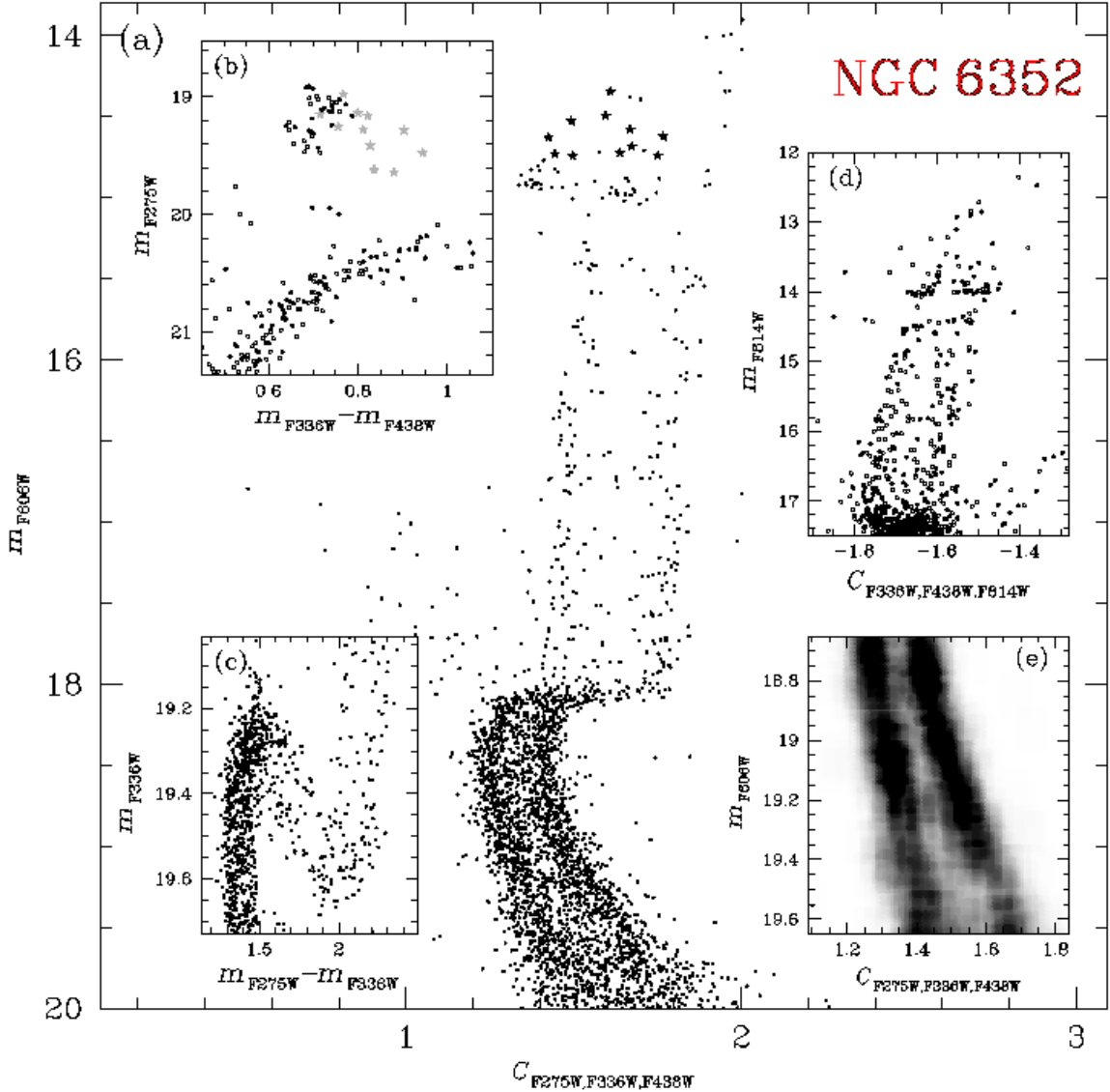}
  \caption{Overview of the main features in the CMDs of
    NGC\,6352. Panel (a) shows the $m_{\rm F606W}$ vs. $C_{\rm
      F275W,F336W,F438W}$ pseudo-CMD, in which the split of all the sequences
    is clear (starred points represent the AGB stars); panel (b) shows
    the region around the HB in the $m_{\rm F275W}$ vs.  $m_{\rm
      F336W}-m_{\rm F438W}$ CMD; panel (c) highlights the SGB split in
    the $m_{\rm F336W}$ vs.  $m_{\rm F275W}-m_{\rm F336W}$ diagram;
    panel (d) shows the RGB region in the $m_{\rm F814W}$ vs. $C_{\rm
      F336W, F438W, F814W}$ pseudo-CMD. 
    The Hess diagram of the MS stars shown in panel (a) is plotted in panel (e) and highlights the two distinct sequences.
  }
  \label{cmds}
\end{figure*}

Figure~\ref{cmds}a shows the $m_{\rm F606W}$ vs. $C_{\rm
  F275W,F336W,F438W}$=($m_{\rm F275W}$$-$$m_{\rm F336W}$)$-$($m_{\rm
  F336W}$$-$$m_{\rm F438W}$) diagram for proper-motion-selected
cluster members.  Asymptotic Giant Branch (AGB), stars selected from
the $m_{\rm F336W}$ vs.\,$m_{\rm F336W}$-$m_{\rm F814W}$ CMD, are
represented with starred symbols.  Two distinct sequences are clearly
visible and can be continuously followed from the main sequence (MS),
the sub-giant branch (SGB), the red giant branch (RGB), to the
HB, and the AGB.  Panels b--e of Fig.~\ref{cmds} show a collection of CMDs for NGC\,6352 and reveal the pattern
of its multiple sequences.

%%__________________________________________________________________
%%__________________________________________________________________

%%__________________________________________________________________
%%__________________________________________________________________

%%%%%%%%%%%%%%%%%%%%%%%%%%%%%%%%%%%%%%%%%%%%%%%%%%%%%%%%%%%%%%%%%%%%%%%%%%%%%%%%%%

\subsection{Multiple stellar populations in NGC 6352}
\label{multpopCMD}
\begin{figure*}
  \centering
  \includegraphics[width=0.95\hsize]{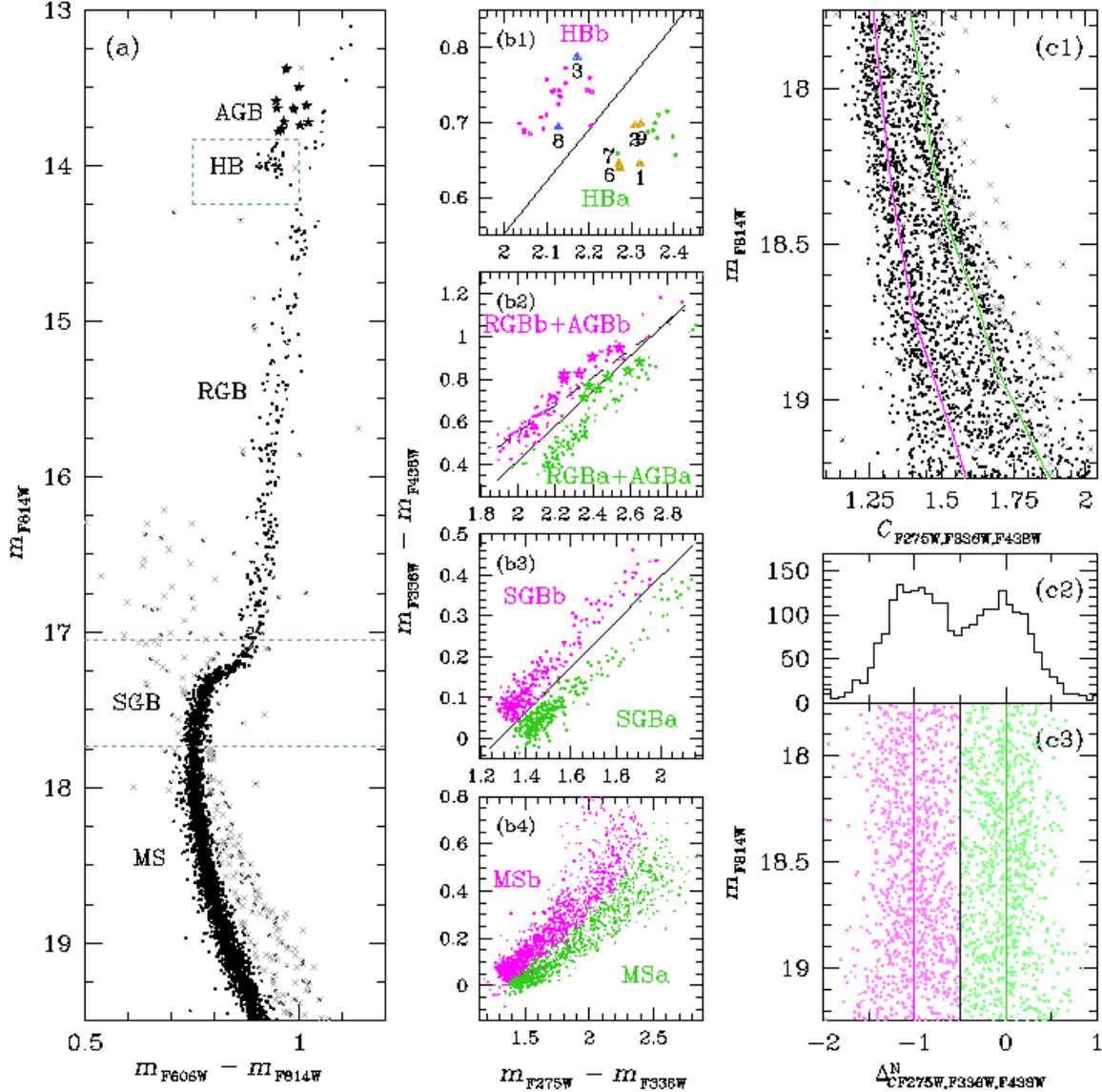}
  \caption{Procedure for the selection of POPa and POPb stars. Panel
    (a) shows the $m_{\rm F814W}$ vs.  $m_{\rm F606W}-m_{\rm F814W}$
    CMD of NGC\,6352: gray crosses are the stars that are possible
    binaries and blue stragglers and that are not used in this
    work. Panels (b) show the $m_{\rm F336W}-m_{\rm F438W}$
    vs. $m_{\rm F275W}-m_{F336W}$ diagrams for MS, SGB, RGB, AGB and
    HB stars of POPa (green) and POPb (magenta). In panel (b1) we also
    plot the stars in common with the spectroscopic catalog of
    \citet{2009A&A...493..913F}. Panels (c) show the procedure used
    for the selection of MSa and MSb stars (see text for details). }
  \label{selpop}
\end{figure*}

In Fig.~\ref{selpop} we use the $m_{\rm F336W}$-$m_{\rm F438W}$
vs.\,$m_{\rm F275W}$-$m_{\rm F336W}$ two-color diagram to separate the
two populations, named population-a (POPa) and population-b (POPb)
hereafter.  Panel (a) of Fig.~\ref{selpop} shows the $m_{\rm F814W}$
vs.\,$m_{\rm F606W}$-$m_{\rm F814W}$ CMD, corrected for differential
reddening and only for cluster-member stars, selected using proper
motions. The green dotted lines identify the four regions in the CMD
which include MS, SGB, RGB and HB stars.  Again, AGB stars are plotted
using starred symbols.  All stars that, on the basis of their position
on the CMD, are possible binaries, blue stragglers or survived field
stars have been excluded from the following analysis and plotted as
gray crosses.

Our recent papers demonstrated how two-color and color-magnitude
diagrams made with appropriate combination of far-ultraviolet,
ultraviolet and blue magnitudes represent a very efficient tool to
identify multiple stellar populations in a GC
(\citealt{2012ApJ...745...27M,2012ApJ...744...58M,2013ApJ...767..120M};
Paper\,I).  Panels (b1), (b2), (b3), and (b4) of Fig.~\ref{selpop}
show the $m_{\rm F336W}$-$m_{\rm F438W}$ vs.\,$m_{\rm F275W}$-$m_{\rm
  F336W}$ two-color diagrams for the four 
evolutionary sequences highlighted in the
left-panel CMD. Two sequences are clearly visible in each diagram.  We
drew by hand a straight continuous line to separate the two groups of
POPa and POPb stars in the SGB, RGB + AGB, and HB, and colored them
green and magenta, respectively. The two sequences of AGBa and AGBb
stars are separated by the black dashed line in panel (b2).  In order
to separate the two MSs we used the $m_{\rm F814W}$ vs.\,$C_{\rm
  F275W,F336W,F438W}$ diagram plotted in panel (c1) of
Fig.~\ref{selpop}. Indeed, in this pseudo-CMD the double MS of
NGC\,6352 is better distinguishable than in the $m_{\rm
  F336W}$-$m_{\rm F438W}$ vs.\,$m_{\rm F275W}$-$m_{\rm F336W}$ plane.

%%__________________________________________________________________
%%__________________________________________________________________

The green and the magenta fiducial lines, superimposed on the MS, are
the fiducials of the MSa and the MSb, respectively, and have been
obtained by using the following procedure. As a first guess, we have
selected by eye a sample of MSa and MSb stars and derived for each of
them a fiducial line by fitting a spline through the median values of
$C_{\rm F275W,F336W,F438W}$ obtained in successive short intervals of
magnitude. We iterated this step with a sigma-clipping procedure.
Then, as in  \citet[Paper\,II]{2015MNRAS.447..927M} , we verticalized the MS in
such a way that the green and magenta fiducials translate into
vertical lines with abscissa $\Delta_{C_{\rm F275W,F336W,F438W}}^{\rm
  N}$=0 and $-1$, respectively. The histogram of the distribution in
$\Delta_{C_{\rm F275W,F336W,F438W}}^{\rm N}$ plotted in panel (c2) is
clearly bimodal. We considered stars with $\Delta_{C_{\rm
    F275W,F336W,F438W}}^{\rm N}>-$0.5 as MSa members and the remaining
MS stars as MSb objects. Panel (c3) shows the $m_{\rm F814W}$
vs. $\Delta_{C_{\rm F275W,F336W,F438W}}^{\rm N}$ verticalized diagram
for the MS stars.  We colored MSa and MSb stars in green and magenta,
respectively. These colors will be consistently used hereafter.

%%%%%%%%%%%%%%%%%%%%%%%%%%%%%%%%%%%%%%%%%%%%%%%%%%%%%%%%%%%%%%%%%%%%%%%%%%%%%%%%%%%%%%%%%%%%%%%

\subsection{The chemical composition of the HB stars} 

Typically, multiple sequences along the CMD correspond to distinct
stellar populations with different content of helium and light
elements (see, e.g., \citealt{2008A&A...490..625M, 2008ApJ...672L..29Y}; Paper\,II).
\citet{2009A&A...493..913F} measured chemical abundances for nine HB
stars of NGC\,6352 from high signal-to-noise UVES@VLT spectra.  They
have confirmed that NGC\,6352 is a metal-rich GC
([Fe/H]=$-$0.55$\pm$0.03), and is enhanced in $\alpha$ elements
by [$\alpha$/Fe]$\sim$0.2 dex.  Feltzing and
collaborators also detected significant star-to-star sodium variation
in close analogy with what  is observed in most 
Milky-Way GCs (e.g., \citealt{1993AJ....106.1490K,
  2004ARA&A..42..385G}).

In order to investigate the chemical content of POPa and POPb, we have
exploited the spectroscopic results by \citet{2009A&A...493..913F}.
Our photometric catalog includes seven stars studied by
\citet{2009A&A...493..913F}. Five of them belong to the HBa and are
Na-poor ([Na/H]$\leq -0.38$). The other two are Na-rich ([Na/H]$\geq
-0.27$) and are HBb members.  The two groups of Na-rich and Na-poor
stars both have the same mean metallicity [Fe/H]$=-0.55\pm 0.02$.  In
panel (b1) of Fig.~\ref{selpop} we plotted these HB stars as triangles
with the corresponding ID number adopted by
\citet{2009A&A...493..913F}: orange triangles are for Na-poor stars,
cyan triangles show the Na-rich ones.
%%%%%%%%%%%%%%%%%%%%%%%%%%%%%%%%%%%%%%%%%%%%%%%%%%%%%%%%%%%%%%%%%%%%%%%%%%

\section{Multi-wavelength view of multiple populations}
\label{sec:multiw}
\begin{figure*}
  \centering
  \includegraphics[scale=0.9]{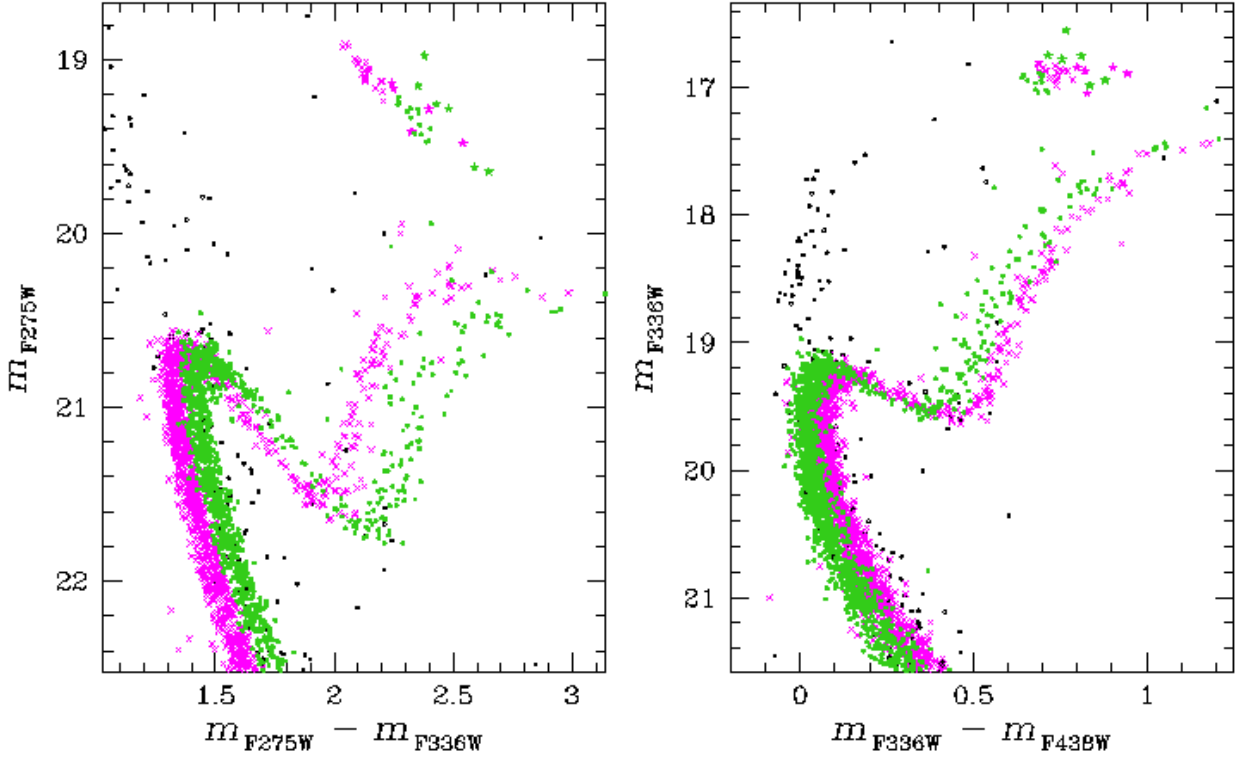}
  \caption{The $m_{\rm F275W}$ vs. $m_{\rm F275W}$-$m_{\rm F336W}$
    (left panel) and $m_{\rm F336W}$ vs. $m_{\rm F336W}$-$m_{\rm
      F438W}$ (right panel) CMDs. In the first CMD, POPa stars
    (green dots) are 
redder than POPb stars (magenta crosses),
    while in the $m_{\rm F336W}$ vs. $m_{\rm F336W}$-$m_{\rm
      F438W}$ CMD POPa 
stars have, on average, bluer colors than POPb stars.
} \label{2pop}
\end{figure*}

\begin{figure*}
  \centering
  \includegraphics[scale=0.9]{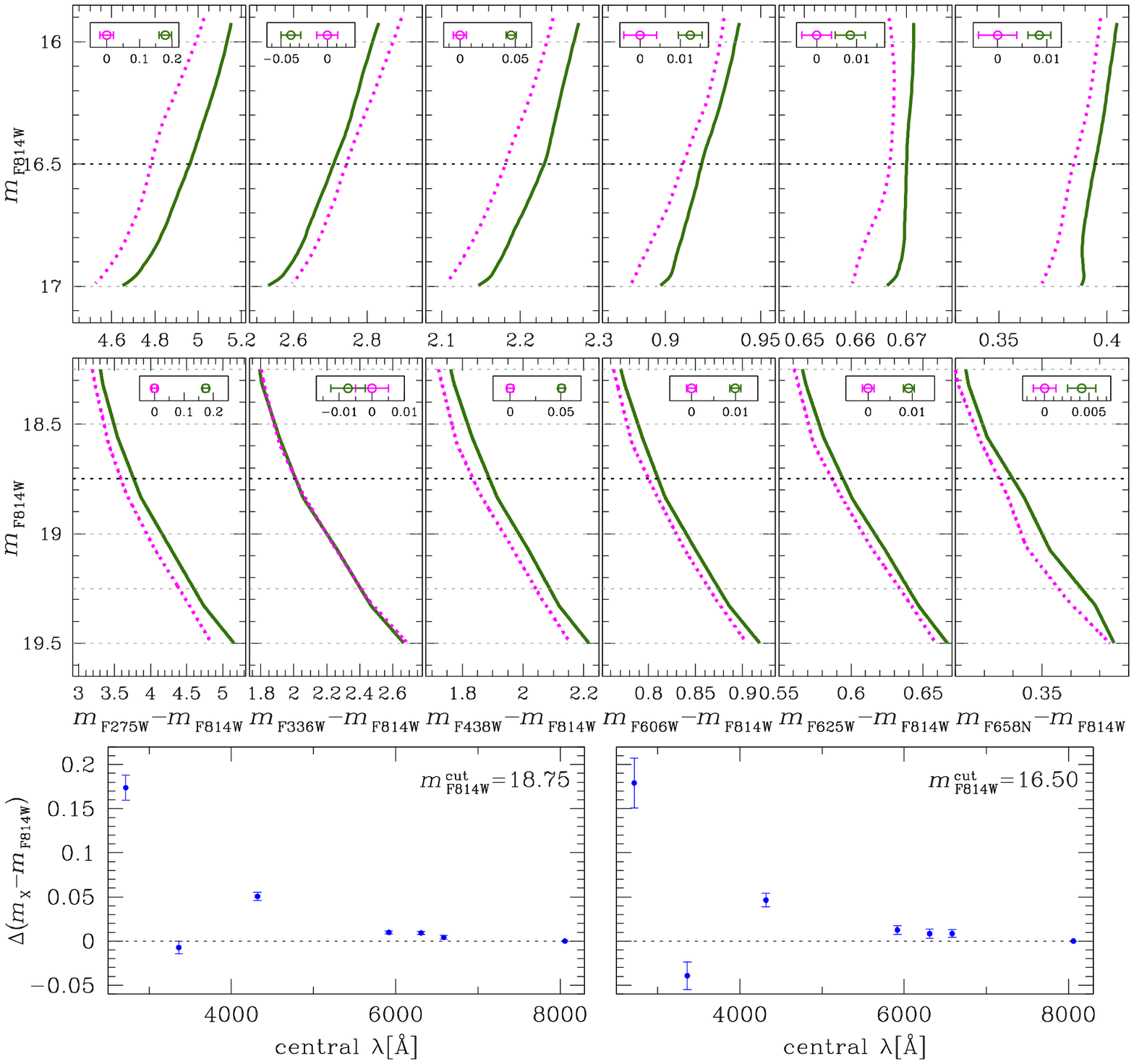}
  \caption{{\it Top Panels:} RGB fiducial lines for POPa (green solid
    lines) and POPb (magenta dotted lines) in six $m_{\rm F814W}$
    vs. $m_{\rm X}-m_{\rm F814W}$ CMDs (where X$=$F275W, F336W, F438W,
    F606W, F625W, F658N); the panel insets show the color distance
    between the two fiducials at $m_{\rm F814W}^{\rm CUT}=16.5$. {\it
      Middle Panels:} same as top panels, but for the MS; the color
    difference indicated in the inset is measured at $m_{\rm F814W}^{\rm CUT}=18.75$.  {\it
      Bottom panels:} $m_{\rm X}-m_{\rm F814W}$ color distance between
    MSa and MSb at $m_{\rm F814W}^{\rm CUT}=18.75$ ({\it left}) and
    between RGBa and RGBb at $m_{\rm F814W}^{\rm CUT}=16.50$ ({\it
      right}) as a function of the central wavelength of the X
    filter.}
  \label{helium}
\end{figure*}

 In the previous section we used the $m_{\rm F814W}$ vs.\,$C_{\rm
   F275W,F336W,F438W}$ pseudo-CMD and the $m_{\rm F336W}$-$m_{\rm
   F438W}$ vs.\,$m_{\rm F275W}$-$m_{\rm F336W}$ two-color diagram,
 where multiple sequences are clearly distinguishable, in order to
 select the members of the two stellar populations
 along the MS, the SGB, the RGB, and the HB of NGC\,6352.
We can now combine all different bands in order to study the behavior
(e.g. color differences) of the two populations in all possible CMDs
we can construct with our data.  An example is shown in
Fig.~\ref{2pop}. Note that POPa stars are redder than POPb ones in
the $m_{\rm F275W}$ vs. $m_{\rm F275W}$-$m_{\rm F336W}$ CMD, while
they are bluer in the $m_{\rm F336W}$ vs. $m_{\rm F336W}$-$m_{\rm
  F438W}$ CMD, as in the case of 47~Tuc \citep{2012ApJ...744...58M}.
The two upper panels of Fig.~\ref{helium} show the $m_{\rm F814W}$
vs.\,$m_{\rm X}-m_{\rm F814W}$ fiducial lines for the RGBs (top
panels) and MSs (middle panels) of the two populations, where X=F275W,
F336W, F438W, F606W, F625W, and F658N.

%%__________________________________________________________________
%%__________________________________________________________________

%%__________________________________________________________________
%%__________________________________________________________________

POPa is redder than POPb in most of the CMDs of Fig.~\ref{helium}, in
close analogy with what  is observed in most GCs
(\citealt{2010AJ....140..631B, 2012ApJ...745...27M,
  2012ApJ...744...58M, 2013ApJ...767..120M}).  In visual and
ultraviolet the $m_{\rm X}-m_{\rm F814W}$ color separation between
POPa and POPb increases with the color baseline, and is maximum for
$m_{\rm F275W}-m_{\rm F814W}$.  The fiducial of the two populations
are almost coincident in $m_{\rm F336W}$-$m_{\rm F814W}$, where POPa
is slightly bluer than POPb.

The procedure used to determine each fiducial line is based on the
naive estimator \citep{1986desd.book.....S}. First, we defined a
series of $N$ $m_{\rm F814W}$ magnitude intervals with a given width
($w$). Magnitude intervals are defined over a grid of $N$ points
separated by step of fixed magnitude ($s$). We calculated the median
magnitude $m_{\rm F814W, i}$ and color ${\rm (}m_{\rm X}-m_{\rm
  F814W}{\rm )}_{\rm i}$ of stars within each magnitude interval
$i=1,...,N$.  These median points have been smoothed using the
smoothing technique of boxcar averaging, in which each point of a
vector is replaced with the average of the $M$ adjacent
points. Finally, the smoothed median points were interpolated with a
cubic spline.

The choice of $w$ is the result of a compromise. On one hand, we
require small magnitude intervals to account for the detailed
structure of the fiducial line. On the other hand, we need a large
width to include in the magnitude interval a large number of stars for
a statistically significant measurement of the median color and
magnitude.  We used different values of $w$ for stars with different
luminosity. Specifically, we assumed $w=0.5$ for RGB stars with $15.00
\leq m_{\rm F814W}<17.00$, $w=0.25$ for SGB stars with $17.00 \leq
m_{\rm F814W}<17.75$, and $w=0.05$ for MS stars with $17.75 \leq
m_{\rm F814W}<19.00$. In all the cases we used $s=w/3$.
  
%%%%%%%%%%%%%%%%%%%%%%%%%%%%%%%%%%%%%%%%
In order to estimate the error associated to the colors ($\sigma_{\rm
  fidcol}$) and magnitudes ($\sigma_{\rm fidmag}$) of each observed
fiducial we used the following procedure.  For each stellar population
selected in Fig.~\ref{selpop} we assigned a subsample of artificial
stars (ASs) and randomly extracted a star from it by following the
recipe in \citet{2009A&A...497..755M}. The AS subsample consists of
all the ASs with similar magnitudes (within 0.3 $m_{\rm F814W}$
magnitudes) and radial distances (less than 100 pixels from the
observed star). This method produces a catalog of simulated stars with
almost the same luminosity and radial distribution of the observed
catalog. We applied the procedure described above to estimate
the fiducial line of the sample of AS stars, and calculated the
difference between the fiducial of the ASs and the real stars. We
repeated this step 100 times.  The color error associated to each
point of the observed fiducial is calculated as the $68.27^{\rm th}$
percentile of this distribution.

\section{The helium abundance of the two stellar populations}
\label{sec:helium}
The helium abundance of a stellar population is a fundamental
ingredient to understand its evolution, and to shed light on the
chemical-enrichment and the star-formation history in a GC.  In
addition, accurate helium determination is crucial to estimate several
parameters of the host star cluster, like age, mass, and mass
function.

A direct spectroscopic determination of
the He content in GCs is only feasible in rare cases and for a tiny
subset of stars (e.g., \citealt{ 2003ApJS..149...67B,
  2009A&A...499..755V, 2011ApJ...728..155D, 2011A&A...531A..35P,
  2014MNRAS.437.1609M}). On the other hand, the method based on
multi-wavelength photometry of multiple sequences that we developed
can be applied to all clusters and provides a reliable estimate of the
He abundance differences among the different populations and can
  reach internal errors smaller then spectroscopic methods (see
\citealt{2012ApJ...744...58M}; Paper\,II and references therein).
In the few cases where helium has been inferred from both
  spectroscopy of HB stars and photometry, the results from the two
  techniques are in fairly agreement. In the case of NGC\,2808,
  \citet{2014MNRAS.437.1609M} derived an helium abundance $Y = 0.34
  \pm 0.01 \pm 0.05$ (internal plus systematic uncertainty) from
  spectroscopy of HB stars slightly bluer than the RR Lyrae
  instability strip.  This result is consistent with the helium
  abundance inferred from the analysis of multiple MSs, where the
  middle MS (which would be associated to the stars analyzed by Marino
  and collaborators), has $Y\sim0.32$
  (\citealt{2007ApJ...661L..53P,2012A&A...537A..77M}).  Similarly in
  the case of NGC\,6397, both photometry
  (\citealt{2010A&A...511A..70D,2012ApJ...745...27M}) and spectroscopy
  (\citealt{2014ApJ...786...14M}) conclude that second-population stars are
  slightly enhanced in $Y$ by $\Delta Y\sim0.01$ dex. In the case of M\,4,
  \citet{2012ApJ...748...62V} found that the second population is
  enhanced in helium by ~0.04-0.05\,dex with respect to the first one. This
  result is in apparent disagreement with results by
  \citet{2015A&A...573A..70N} who inferred an helium difference of
  $0.020\pm0.008$ dex between the two MSs of this clusters. However,
  it should be noted that the discrepancy could be due to NLTE that
  affect the helium line at $5875.6\,\AA$ (see e.g. Marino et
  al. 2014, Sect. 4 for the case of NGC\,2808).  

Helium has been also inferred in RGB stars from near-infrared
chromospheric transition of He\,I at 10\,830$\AA$ in NGC\,2808 and
$\omega$\,Centauri. The spectra suggest helium abundances of $Y<$0.22
and $Y=$0.39-0.44 ($\Delta Y \ge 0.17$) for the two analyzed stars in
$\omega$\,Centauri (\citealt{2013ApJ...773L..28D}), while Na-rich and
Na-poor stars of NGC\,2808 analyzed by \citet{2011A&A...531A..35P}
differ in helium by $\Delta Y>$0.17, with the sodium-rich star being
also helium rich.  Such helium variation are larger than those
obtained from the analysis of multiple sequences in the CMD, that are
$\sim0.07$ for NGC\,2808
(\citealt{2005ApJ...631..868D,2007ApJ...661L..53P,2012A&A...537A..77M})
and $\sim0.13$ for $\omega$\,Centauri (\citealt{2012AJ....144....5K}).

Figure~\ref{helium} illustrates the procedure to infer the helium
content of the stellar populations in NGC\,6352. Upper panels show a
collection of $m_{\rm F814W}$ vs.\,$m_{\rm X}-m_{\rm F814W}$ fiducial
lines, where X$=$F275W, F336W, F438W, F606W, F625W and F658N. We have
used green and magenta continuous lines to plot the fiducials along the
RGB (top panels) and the MS (middle panels) of the two populations.
We have calculated the color difference between the green and the
magenta fiducials at the reference magnitudes indicated by the black
dashed lines at $m_{\rm F814W}^{\rm CUT}$=18.75 for the MS, and
$m_{\rm F814W}^{\rm CUT}$=16.50 for the RGB.

Results are shown in the lower panels of Fig.~\ref{helium}. We found
that $\Delta$($m_{\rm X}-m_{\rm F814W}$) has negative values only for
X=F336W.  When the other filters are used, the
$m_{\rm X}-m_{\rm F814W}$ color separation between the two RGBs and
MSs is maximum for X=$m_{\rm F275W}$ and decreases towards redder
wavelengths of the X filter.

In order to estimate the effective temperatures ($T_{\rm eff}$) and
gravities ($\log{g}$) of MS and RGB stars with $m_{\rm
  F814W}$=$m_{\rm F814W}^{\rm CUT}$ we used the isochrones by
\citet{2008ApJS..178...89D} that best match the CMD.

%%%%%%%%%%%%%%%%%%%%%%%%%%
\begin{table*}
  \begin{tabular}{rrrrrrr}
    \hline
    \hline
        {\bf $m^{\rm CUT}_{\rm F814W}$} & {\bf $\Delta m_{\rm F275W,F814W}$}  & {\bf $\Delta m_{\rm F336W,F814W}$} & {\bf $\Delta m_{\rm F438W,F814W}$}  & {\bf $\Delta m_{\rm F606W,F814W}$}  & {\bf $\Delta m_{\rm F625W,F814W}$}  & {\bf $\Delta m_{\rm F658N,F814W}$} \\ 
        \hline
        19.50   &   0.300$\pm$0.028   &  -0.016$\pm$0.017   &   0.049$\pm$0.011   &   0.012$\pm$0.003   &   0.005$\pm$0.004   &   0.001$\pm$0.004   \\
        19.25   &   0.234$\pm$0.026   &  -0.009$\pm$0.012   &   0.045$\pm$0.007   &   0.011$\pm$0.002   &   0.006$\pm$0.003   &   0.005$\pm$0.003   \\
        19.00   &   0.204$\pm$0.019   &  -0.003$\pm$0.009   &   0.048$\pm$0.006   &   0.012$\pm$0.002   &   0.009$\pm$0.002   &   0.005$\pm$0.003   \\
        18.75   &   0.174$\pm$0.014   &  -0.007$\pm$0.007   &   0.051$\pm$0.005   &   0.010$\pm$0.002   &   0.009$\pm$0.002   &   0.004$\pm$0.002   \\
        18.50   &   0.128$\pm$0.011   &  -0.011$\pm$0.006   &   0.045$\pm$0.004   &   0.007$\pm$0.002   &   0.005$\pm$0.002   &   0.001$\pm$0.002   \\
        18.25   &   0.111$\pm$0.008   &  -0.008$\pm$0.004   &   0.038$\pm$0.003   &   0.008$\pm$0.001   &   0.005$\pm$0.001   &   0.001$\pm$0.002   \\
        17.00   &   0.168$\pm$0.038   &  -0.056$\pm$0.030   &   0.052$\pm$0.018   &   0.016$\pm$0.008   &   0.016$\pm$0.008   &   0.022$\pm$0.009   \\
        16.50   &   0.179$\pm$0.028   &  -0.039$\pm$0.016   &   0.046$\pm$0.008   &   0.013$\pm$0.005   &   0.008$\pm$0.005   &   0.009$\pm$0.005   \\
        16.00   &   0.158$\pm$0.030   &  -0.033$\pm$0.020   &   0.044$\pm$0.008   &   0.011$\pm$0.005   &   0.010$\pm$0.007   &   0.010$\pm$0.006   \\
        \hline
  \end{tabular}
  \caption{Color difference $\Delta m_{\rm X,F814W}$ at different $m^{\rm CUT}_{\rm F814W}$. \label{tab1}}
\end{table*}

%%%%%%%%%%%%%%%%%%%%%%%%%%%%%%%%%%%%%%%%%%%%%%%%%%%%%%%%%%%%%%%%%%%%%%%%%%%%%%%%%%%%%%%%%%

Note that we are mainly interested in relative ages rather than in
absolute ones. Indeed, as demonstrated in next section, accurate
relative ages are feasible using our dataset.
Here, we just need reference models to calculate the color
differences of the two populations. As discussed in
Section~\ref{sec:age},
the best match with observations could be obtained using isochrones
with [Fe/H]=$-0.67$, in fair agreement with the value inferred from
high-resolution spectroscopy by \citet{1997A&AS..121...95C},
[$\alpha$/Fe]=0.4, and an age of 13.1 Gyr [adopting $(m-M)_V=14.43$,
  an $E(B-V)=0.22$, \citealt{1996AJ....112.1487H}, 2010 edition].  We
assumed $Y=0.256$ for POPa, and different helium abundances with $Y$
ranging from 0.256 to 0.330, in steps of 0.001, for POPb. Since there
are no measurements of C, N, O abundance for NGC\,6352 stars, we
assumed for POPa and POPb the same C, N, O content as the two main
stellar populations of 47\,Tuc. Specifically, we used for POPa
[C/Fe]=-0.20, [N/Fe]=0.20, [O/Fe]=0.40, and assumed that POPb stars
are depleted in carbon and oxygen by 0.20 and 0.15 dex, respectively
and nitrogen enhanced by 0.7 dex with respect to POPa stars.

We used the ATLAS12 code (\citealt{1993sssp.book.....K,
  2004MSAIS...5...93S}) to calculate atmospheric models for the MS and
RGB stars with $m_{\rm F814W}$=$m_{\rm F814W}^{\rm CUT}$, while
synthetic spectra have been generated with SYNTHE
(\citealt{1981SAOSR.391.....K}) with a resolution of $R$=600 from 2000
to 10000\AA.

Synthetic spectra of POPa and POPb stars have been integrated over the
transmission curves of the WFC3/UVIS and the ACS/WFC filters used in
this paper in order to determine synthetic colors. The colors from synthetic
spectra with different $Y$ have been compared with observations in order to
estimate the helium abundance of POPb.

The best fit between observed and theoretical colors at $m_{\rm
  F814W}^{\rm CUT}=$18.75 can be obtained assuming POPb enhanced in
helium by $\Delta$Y=0.033~dex.  We have repeated this procedure for
$m_{\rm F814W}^{\rm CUT}=$18.25, 18.50, 19.00, 19.25, and 19.50 for
the MS, and $m_{\rm F814W}^{\rm CUT}=$16.00, 16.50 and 17.00 for the
RGB. The results are listed in Table~\ref{tab1}. The values of $T_{\rm
  eff}$, log($g$), and $\Delta$Y that provide the best fit to the
observed color differences are listed in Table~\ref{tab2} for each
value of $m_{\rm F814W}^{\rm CUT}$. On average, we have
$\Delta$Y=0.029$\pm$0.006, where the error represent the 68.27th
percentile of the distribution of the sorted residuals from the mean
value.  To estimate $\Delta$Y we only used visual bands (F606W, F625W,
F658N, and F814W), because they are not affected by light-elements
variations. For this reason, the final result is not conditioned by
the assumptions on the C, N, and O abundances.  
 In the case where C, N, and O abundances were known, the final
  value of $\Delta Y$, obtained using all the photometric bands, would
  not change. As a test, we repeated the procedure described above by
  assuming different values of C, N, and O: we found the same result.

%%%%%%%%%%%%%%%%%%%%%%%%%%%

\begin{table}
  \small
  \begin{tabular}{rrrrrc}
    \hline
    \hline
        {\bf $m^{\rm CUT}_{\rm F814W}$} & {\bf $T_{\rm EFF,POPa}$}  & {\bf $T_{\rm EFF, POPb}$} & {\bf $\log{g}_{\rm POPa}$}  & {\bf $\log{g}_{\rm POPb}$}  & $\Delta Y$\\ 
        \hline
        19.50   &  3673   &    3677    &   2.51   &    2.53   &   0.018 \\
        19.25   &  3678   &    3684    &   2.64   &    2.69   &   0.023 \\
        19.00   &  3687   &    3692    &   2.86   &    2.91   &   0.030 \\
        18.75   &  3693   &    3698    &   3.06   &    3.10   &   0.033 \\
        18.50   &  3698   &    3702    &   3.23   &    3.26   &   0.024 \\
        18.25   &  3700   &    3705    &   3.37   &    3.40   &   0.031 \\
        17.00   &  3766   &    3767    &   4.25   &    4.26   &   0.042 \\
        16.50   &  3764   &    3766    &   4.29   &    4.30   &   0.026 \\
        16.00   &  3754   &    3756    &   4.40   &    4.41   &   0.030 \\
        \hline
                &         &            &          &           & 0.029$\pm$0.006 \\
        \hline
  \end{tabular}
  \caption{Parameters used to simulate synthetic spectra of POPa and
    POPb stars and estimation of helium difference between the two
    populations for different $m^{\rm CUT}_{\rm F814W}$. \label{tab2}}
\end{table}

%%%%%%%%%%%%%%%%%%%%%%%%%%%%%%%%%%%%%%%%%%%%%%%%%%%%%%%%%

\section{Relative ages of the two stellar populations}
\label{sec:age}

The relative ages of the multiple stellar populations hosted by GCs is
an important issue to understand how GCs formed. We can speculate
that the phenomenon of multiple stellar populations in GCs is due to
the presence of different generations of stars, formed in different
epochs: a first stellar generation (FG) characterized by primordial
helium and chemical composition similar to that of field stars with
the same metallicity, and a helium-enhanced second generation (SG),
characterized by stars depleted in C and O and enhanced in N and Na,
born from material processed at high temperature by FG stars. If we
exclude the ``anomalous'' GCs, such as $\omega$~Cen
(\citealt{1995ApJ...441L..81N}), M\,22
(\citealt{2011A&A...532A...8M}), and M\,2
(\citealt{2014MNRAS.441.3396Y}, Paper\,II), stars of different
populations in ``normal'' GCs show negligible differences in
metallicity (\citealt{2009A&A...508..695C}). Because of this, we can
exclude supernova ejecta as possible polluters of the material from
which SG stars formed.

Current scenarios for the formation of multiple stellar
populations in GCs predict different timescales. The AGB scenario
(\citealt{2002A&A...395...69D}) predicts that envelopes of
intermediate mass AGB stars are the cause of pollution. 
In models aimed at reproducing the observed chemical patterns in
``normal'' globular clusters with AGB ejecta, the second generation star
formation epoch extends between about 40 and 90--100 Myr
\citep{2010MNRAS.407..854D,2012MNRAS.423.1521D}.
Another suggested mechanism is the formation of stars from material
ejected by Fast-Rotating Massive Stars (FRMS) during the phase of core
H-burning (\citealt{2007A&A...464.1029D}) or massive binaries
(\citealt{2013MNRAS.436.2398B, 2014A&A...566A.109S}). 
In these cases, the timescale for the formation of the second
generation must be of the order of a few Myrs or even
less\,(\citealt{2013MNRAS.436.2398B,2014A&A...566A.109S,2014MNRAS.443.3302D}).

In this work we present a method to set upper limits on the difference
in age between the two populations of NGC\,6352. For the first time,
we estimate the relative age of two stellar populations in a normal
GC, i.e. with no internal variation in [Fe/H].  Note that we are
interested in measuring relative ages. The corresponding absolute ages
should be only regarded as indicative (depending on the adopted
distance modulus and reddening, as well as the reference model).
%%%%%%%%%%%%%%%

We consider a set of isochrones characterized by [Fe/H]= $-0.67$ and
[$\alpha$/Fe]=0.4, that are the values that best fit all the stars in
the $m_{\rm F814W}$ versus $m_{\rm F606W}-m_{\rm F814W}$, as
explained in Sect.~\ref{sec:helium}.

\begin{figure*}
  \centering
  \includegraphics[width=0.95\hsize]{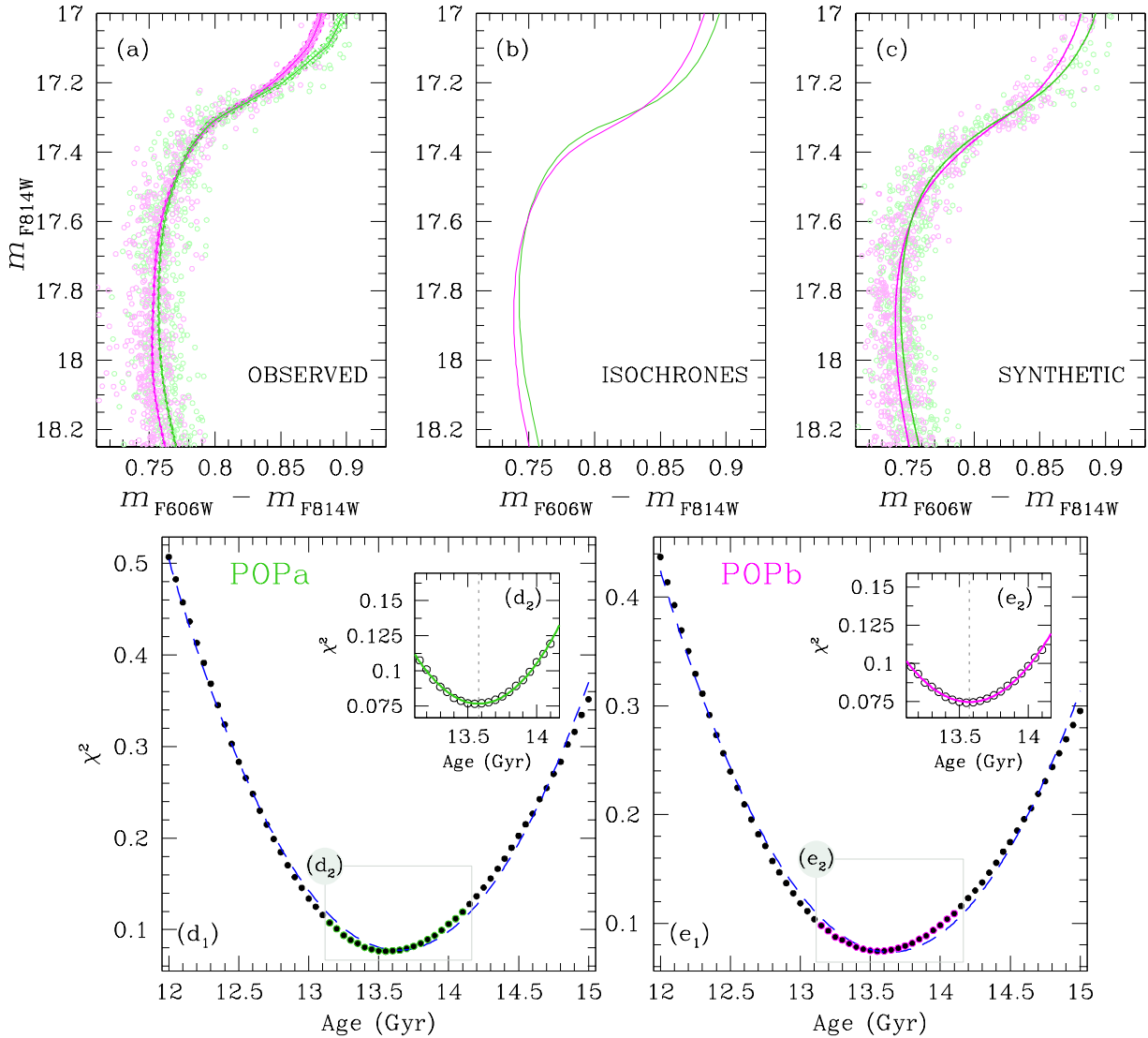}
  \caption{Procedure used to obtain the relative age of the two
    populations. In panel (a) the observed $m_{\rm F814W}$ vs. $m_{\rm
      F606W}-m_{\rm F814W}$ CMD is shown. In green and magenta are the
    fiducial lines of the POPa and POPb, respectively.  Panel (b)
    shows the best-fit isochrones for the two populations, while in
    panel (c) we plot the synthetic CMD and the fiducial lines built
    using the isochrones calculated for the two population.  Panels
    (d) and (e) illustrate the procedure adopted to derive the age of
    POPa and POPb, respectively. In panels (d$_1$) and (e$_1$), filled
    circles indicate the values of $\chi^2$ for different ages while
    the dashed lines are the first-guess second-order polynomial
    fit. The insets (d$_2$) and (e$_2$) show the best-fit second-order
    polynomial for POPa (green) and POPb (magenta, see text for
    details).}
  \label{age}
\end{figure*}

We also used the $m_{\rm F814W}$ versus $m_{\rm F606W}-m_{\rm F814W}$
CMD for computing the relative ages between the two populations,
because in these filters the effect of light elements variation is not
significant (\citealt{2011A&A...534A...9S,2012ApJ...744...58M}). The
MS turn-off is the classical age indicator of a simple stellar
population. For this reason, we considered a region of the CMD around the MS
turn-off of the two populations [panel (a) of Fig.~\ref{age}]. We
obtained fiducials and the associated errors in color and magnitudes (colored
regions between dashed lines) for POPa (green) and POPb (magenta) using
the same procedure described in Sect.~\ref{sec:multiw}. 

In order to compute the relative age between the two populations, we
treated POPa and POPb as simple stellar populations and measured their
ages independently. The age of each population is estimated by
comparing the observed $m_{\rm F814W}$ versus 
$m_{\rm F606W}-m_{\rm F814W}$
CMD with a grid of synthetic CMDs with the same [Fe/H],
[$\alpha$/Fe] and $Y$, and ages that run from 12000\,Myr to
15000\,Myr, in step of 50\,Myr.

For each isochrone a synthetic CMD was built in this way: we obtained
a first guess synthetic CMD by interpolating the isochrone on the
observed magnitudes, in such a way that 
to each $m_{\rm F814W}$ magnitude in the catalog of real stars
is associated a color $m_{\rm F606W}-m_{\rm F814W}$ on the isochrone
As second step, we broadened this synthetic CMD by adding to the color
and magnitude of each synthetic star a random Gaussian error, with a
dispersion equal to the error of the associated real star. The final
result is a synthetic CMD, characterized by the same number of stars
and a similar luminosity function to that of the observed CMD.

Panels (b) and (c) of Fig.~\ref{age} show an example of isochrones and
corresponding synthetic CMDs.  For each synthetic CMD we computed the
fiducial line using the same procedure as for the observed data. For
each step in age, we built $N=100$ synthetic simple stellar population
CMDs, obtained the fiducial lines and computed the average of all the
100 fiducial lines in order to obtain the final synthetic fiducial line.

%%__________________________________________________________________
%%__________________________________________________________________

We compared the observed fiducial line of each population with the
synthetic ones. This method is more robust than the direct
comparison between the observed CMD and the isochrones, because we
compared observed and theoretical fiducial lines which were computed
in the same way and are affected by the same systematic errors
introduced by the method (as, e.g., the error due to the smoothing).
For each step in age, we calculated the $\chi^2$ of the best fit
between the synthetic and the observed fiducial, in a magnitude
interval $17.00<m_{\rm F814W}<18.2$. We fitted the $\chi^2$ as a
function of the age with a second-order polynomial, in order to find
the age that minimizes the $\chi^2$.  This fit was performed in two
steps: first, we fitted all the points with a parabola to obtain a
first-guess value of the age, $t^{\rm I}$, that minimizes the $\chi^2$
[panels (d$_1$) and (e$_1$) of Fig.~\ref{age}]. Then we considered
only the points inside an interval between $t^{\rm I}-500$\,Myr and
$t^{\rm I}+500$\,Myr and fitted a new second-order polynomial [panels
  (d$_2$) and (e$_2$)] to find the best value that minimizes the
$\chi^2$.
%%%%

We independently run this procedure for POPa and POPb. For POPa we
used a set of isochrones calculated for [Fe/H]$=-0.67$,
[$\alpha$/Fe]$=+0.4$ and $Y=0.256$, while for POPb we used isochrones
with the same metallicity and $\alpha$-enhancement, but enhanced in
helium by $\Delta Y=0.029$ (i.e. with $Y=0.285$) as obtained in
Section~\ref{sec:helium}.
The best fit gives an age of $(13580\pm80)$\,Myr for POPa and
$(13570\pm80)$\,Myr for POPb.  We want to emphasize the fact that
absolute ages are not significant in this context, and that we are
interested only to relative ages. For this reason, hereafter we
provide only relative ages.
The age error is given by $\sigma =
\sqrt{\sigma_{\rm PHOT}^2+\sigma_{\rm FIT}^2}$, where $\sigma_{\rm
  PHOT}$ is the error due to photometric errors and $\sigma_{\rm FIT}$
is the error on the fit of the $\chi^2$ values.

The error due to photometry has been computed by performing the same
analysis as described above and shifting 
the observed fiducial line in color by a quantity equal to $\pm
\sigma_{\rm fidcol}$ and in magnitude by a quantity equal to $\pm
\sigma_{\rm fidmag}$.
To estimate $\sigma_{\rm FIT}$, we used the bootstrap re-sampling of
the data to generate 1000 samples randomly drawn from the original
data sets; for each sample, we calculated the age that minimize the
$\chi^2$ as above described.  Finally we computed the mean age and its
standard deviation, and adopted the latter as the uncertainty in the
$\chi^2$ fit.
In conclusion, the two populations have a difference in age
$\Delta_{\rm AGE} = {\rm AGE_{POPa}-AGE_{POPb}}=(10 \pm 110)$\,Myr.

Figure~\ref{age} shows the whole procedure: in panel (a) we plot
the CMDs of POPa and POPb (light green and magenta) and their fiducial
lines.
In panel (b) the best-fit isochrones are plotted for each population;
these are the isochrones for which the corresponding synthetic
fiducials, shown in panel (c) with the synthetic CMDs, give the minimum
$\chi^2$.  Panels (d) and (e) show the procedure adopted for the fit
of the $\chi^2$ for POPa and POPb respectively. In panels (d$_1$) and
(e$_1$) are shown the first guess fits (blue dashed lines), while in
panels (d$_2$) and (e$_2$) show the final fits (green and magenta
curves for POPa and POPb respectively). The vertical dashed lines
represent the value of the age that minimizes the $\chi^2$.

We performed the same analysis using CMDs with different color bases:
$m_{\rm F625W}-m_{\rm F814W}$ and $m_{\rm F658N}-m_{\rm F814W}$. These
CMDs are characterized by larger photometric errors. We found
$\Delta_{\rm AGE}=80 \pm 140$\,Myr in the case of the $m_{\rm F814W}$
vs.  $m_{\rm F625W}-m_{\rm F814W}$ CMD, and $\Delta_{\rm AGE}=220 \pm
210$\,Myr in the case of the $m_{\rm F814W}$ vs.  $m_{\rm
  F658N}-m_{\rm F814W}$ CMD. The results are consistent with what
found using the $m_{\rm F814W}$ versus $m_{\rm F606W}-m_{\rm F814W}$,
but errors due to photometric uncertainties are larger. For this
reason we decided to carry out further analysis only using the $m_{\rm
  F814W}$ vs.  $m_{\rm F606W}-m_{\rm F814W}$ CMD.

We tested whether our results depend on the adopted magnitude interval
for the fit. We iterated the procedure described above by changing the
starting point of the magnitude interval between $m_{\rm F814W}=17.80$
and $m_{\rm F814W}=18.10$, and the ending point between $m_{\rm
  F814W}=17.00$ and $m_{\rm F814W}=17.30$, with steps of 0.05 mag.
Within 50~Myr, we found no difference in the average age.
Such age difference is consistent with zero within our measurement errors.

We tested our results using a different dataset of isochrones, to
prove that results are independent from adopted models. We considered
BaSTI isochrones (\citealt{2004ApJ...612..168P,2009ApJ...697..275P})
and we performed the same analysis described above. We obtained a
difference in age between the two populations of $(-80\pm 110)$\,Myr,
in agreement with the results obtained previously.

There is some dispersion among the abundance measurements of NGC\,6352
in the literature (see \citealt{2009A&A...493..913F} and discussion
therein).  Therefore, we re-iterated the same analysis using another
set of isochrones with different [Fe/H] and [$\alpha$/Fe]. Following
the spectroscopic results by \citet{2009A&A...493..913F}, we
considered a set of isochrones with [Fe/H]=$-0.55$ and
[$\alpha$/Fe]=+0.2 and performed the same procedure as described
above. We found for the two populations a slightly different relative
age ($\Delta_{\rm AGE} =70 \pm 110$\,Myr).  We thus confirm that the
two populations have a difference in age within $\sim$110~Myr and that
the adopted values of [Fe/H] and [$\alpha$/Fe] do not change our
conclusions.

In addition to the photometric error, we also considered the
uncertainty in $\Delta Y$, that in the previous section we estimated
to be 0.006 dex (see Table~\ref{tab2}).  In order to estimate how this
error affects the measure of $\Delta_{\rm AGE}$, we iterated the same
procedure previously described, using for POPa the same set of
isochrones, and for POPb two additional sets of isochrones with the
same [Fe/H]=$-0.67$ and [$\alpha$/Fe]=+0.4 as for POPa, but with two
different helium enhancements: $\Delta Y=0.029+0.006=0.035$ and
$\Delta Y=0.029-0.006=0.023$.  We found that an error of 0.006 dex in
$\Delta Y$ translates into an uncertainty in $\Delta_{\rm AGE}$ of
$\sigma_{\rm \Delta AGE}(Y)= 40$~Myr.

The AGB and FRMS models predict that the stars of SG have the same
[Fe/H] and [$\alpha$/Fe] of stars of FG, as observed in many GCs. Moreover
the spectroscopic results by \citet{2009A&A...493..913F} confirm that
in NGC\,6352 the stars of the two populations should have the same
metallicity (within 0.02 dex). In any case, we also considered the
cases in which POPb has different [Fe/H] or [$\alpha$/Fe] with
respect to POPa, to study the impact of these variations on
$\Delta_{\rm AGE}$.
As first test, we considered the case in which POPb has metallicity
${\rm [Fe/H]_{POPb}}$ = ${\rm [Fe/H]_{POPa}\pm 0.02\,dex}$, consistent
with the results by \citet{2009A&A...493..913F}. We performed the
procedure described above using for POPa a set of isochrones with
[Fe/H]=$-0.67$, [$\alpha$/Fe]=0.4 and primordial helium, and for POPb
[Fe/H]=$-0.65$, [$\alpha$/Fe]=0.4, and $Y=0.285$. 
We found that the relative age between the two populations is
$\Delta_{\rm AGE}=210$\,Myr.  In the same way, we considered for POPb
a set of isochrones with the same parameters, except the metallicity
that we set at [Fe/H]=$-0.69$.
In this case we found $\Delta_{\rm AGE}=-190$\,Myr Therefore, a
variation on the metallicity of the second population of
$\delta$[Fe/H]$=\pm 0.02$\,dex translates in a variation on
$\Delta_{\rm AGE}$ of $\sigma_{\rm \Delta AGE}({\rm [Fe/H]})=\pm
200$~Myr.

We then considered the impact of the $\alpha$-enhancement variations
on the estimate of $\Delta_{\rm AGE}$. Similarly to what was done already
for the variations in metallicity, we considered a set of isochrones
for POPa with [Fe/H]=$-0.67$, [$\alpha$/Fe]=0.4 and $Y=0.256$, and for
POPb the same [Fe/H], enhanced in helium by $\Delta Y = 0.029$ and
$\delta [\alpha/{\rm Fe}] = \pm 0.02$ with respect to POPa. We found
that a variation of $\delta$[$\alpha$/Fe]$=\pm 0.02$\,dex leads to a
variation on $\Delta_{\rm AGE}$ of $\sigma_{\rm \Delta AGE}({\rm
  [\alpha/Fe]})=\pm 150$~Myr.

It is possible that POPa and POPb have different C+N+O
abundances. Unfortunately, we are not able to directly verify the
effects of C+N+O abundances variations on the estimate of $\Delta_{\rm
  AGE}$. \citet{2012ApJ...746...14M} presented a [Fe/H]-independent
relation between the C+N+O variations and the age of a population
$\partial{\rm AGE}/\partial{\rm [CNO]}\sim -3.3\,{\rm Gyr\,dex^{-1}}$.
If the POPb is CNO-enhanced with respect the POPa, then POPb is
younger compared to the case in which both populations have the same
CNO abundances, and therefore $\Delta_{\rm AGE}$ is larger.  However
we note that the $m_{\rm F814W}$ and $(m_{\rm F606W}-m_{\rm F814W})$
magnitude and color on which our analysis is based are the least
affected by C, N, O variations.

In Table~\ref{tab3} we provide a summary of all the uncertainties on
$\Delta_{\rm AGE}$ introduced by the variation of the considered
parameters.

\begin{table}
\large
\begin{tabular}{rm{3cm} cm{5cm} cm{5cm}}
\hline
\hline
{\bf Parameter} & {\bf Variation} & {\bf $\sigma_{\rm \Delta AGE}$} \\
\hline
Photometry        &                       &  $\pm110$~Myr \\
$\Delta Y$        &  $\pm 0.006$          &  $\mp40$~Myr \\
$[$Fe/H$]$        &  $\pm 0.02$           &  $\pm200$~Myr \\  
$[\alpha/$Fe$]$   &  $\pm 0.02$           &  $\pm150$~Myr \\  
C+N+O         & \multicolumn{2}{c}{$\partial{\rm AGE}/\partial{\rm [CNO]}\sim -3.3\,{\rm  Gyr\,dex^{-1}}$}\\
\hline
\end{tabular}
\caption{Summary of all the uncertainties on $\Delta_{\rm
    AGE}$ \label{tab3}}
\end{table}

\section{Conclusions}
\label{sec:conclusion}
In this paper we have presented an analysis of the CMDs and two-color
diagrams from WFC3/UVIS F275W, F336W and F438W, and ACS/WFC F625W, F658N,
F606W and F814W photometry.

We identified two stellar populations (named POPa and POPb), that,
with appropriate combinations of magnitudes and colors, we could clearly
distinguish on most of the evolutionary sequences of the CMD, from the
MS, to the SGB, RGB, AGB, and HB.

Using a multi-color analysis method we have already applied to half
a dozen of other GCs, we estimated a $\Delta Y = 0.029 \pm 0.006$ between
the two populations.

We also developed a new procedure for evaluating the difference in age
between the two populations hosted by NGC\,6352.  We considered the
filter combinations that are the least affected by light-element
variations. We studied the case in which POPa and POPb have the same
[Fe/H], as inferred from high-resolution spectroscopy by
\citet{2009A&A...493..913F}, but POPb has helium abundance enhanced by
$\Delta Y= 0.029 \pm 0.006$.  We found that the two populations have a
difference in age between the POPa and POPb of $\Delta_{\rm
  AGE}=10\pm110$\,Myr, where the error includes the photometric error
and the error on the fit. The error in $\Delta Y$ leads to an error on
the relative age of $\sigma_{\rm \Delta AGE} = \pm 40$\,Myr, and a
total error of 120\,Myr.  This is the best relative ages between the
two populations we can estimate assuming that the two populations have
the same [Fe/H] and [$\alpha$/Fe].

In the case that POPb has a different [Fe/H] content of $\pm$0.02\,dex
(as suggested by high-resolution spectroscopy by
\citealt{2009A&A...493..913F}), the relative ages would change by $\pm
200$~Myr.
For a difference in [$\alpha$/Fe] of $\pm 0.02$\,dex, we found a
$\Delta_{\rm AGE}$ variation of $\pm150$ Myr.  As shown by
\citet{2012ApJ...746...14M}, if POPb is CNO-enhanced then it must
be younger since the turn-off brightness is the same. Differences in
metallicity between two populations hosted by a normal GC, however,
are not predicted by ``AGB'' and ``FRMS'' models of formation of
multiple stellar populations and are also not observed for this
cluster.

In literature, there are numerous works on the relative age of
multiple stellar populations hosted by anomalous GCs for which a
spread in metallicity is measured. Surely, the most controversial case
is represented by $\omega$\,Centauri. \citet{2008MmSAI..79..342S}
identified four coeval populations with different metallicities and
helium contents, while \citet{2014ApJ...791..107V} identified a large
spread in age for each population and found that the most metal-rich
population also is the oldest one, concluding that $\omega$\,Centauri
is the result of a merger of two different
progenitors. \citet{2012A&A...541A..15M} found that the populations of
M\,22, characterized by different metallicities, are almost coeval
within $\sim 300$\,Myr. \citet{2013ApJ...778L..13L} found that the GC
NGC\,2419 hosts two populations characterized by a large difference in
[Fe/H], $Y$ and with a relative age of about
2\,Gyr. \citet{2011ApJ...733L..45R} compared models and CMDs of
NGC\,288 to put constraints on the second stellar population hosted by
this GC. They found that, in order to properly reproduce the observed
CMDs, the second generation must be moderately metal enhanced by 0.16
dex, helium-enhanced by 0.03 dex, and younger by 1.5~Gyr than the
first stellar generation.

The work we present in this paper is different from previous studies
on the relative ages of multiple stellar populations in GCs, in the
sense that this is the first attempt to measure relative ages within a
multiple population GC with no observed signature of [M/H] dispersion.
Assuming no difference in [Fe/H], [$\alpha$/Fe], and C+N+O content,
the two populations of NGC6352 have a $1\,\sigma$ spread in age
$\sim120$ Myr.  A combination of small differences in [Fe/H] and
[$\alpha$/Fe] of 0.02 dex would rise the total uncertainty on the
relative age to $\sim$280 Myr (within which the two populations are
still coeval).

\section*{Acknowledgements}
DN is supported by a grant ``Borsa di studio per l'estero, bando
2013'' awarded by ``Fondazione Ing.\,Aldo Gini'' in Padova (Italy).
SC, SO, and GP acknowledge partial support by PRIN-INAF 2014.  SO and
GP acknowledge partial support by Progetto di Ateneo (Universit\`a di
Padova) 2014. APM acknowledges support by the Australian Research
Council through Discovery Early Career Researcher Award DE150101816.

%__________________________________________________________________
\bibliographystyle{mn2e} 
\bibliography{biblio}
\bsp

\label{lastpage}

\end{document}